\documentclass[a4paper,11pt]{article}

\usepackage{amssymb}
\usepackage{graphicx}

\nocite{*}

\begin{document}

\title{Persistent currents and magnetic flux trapping
in fragments of carbon deposits containing multiwalled nanotubes}

\author{V.I.~Tsebro\thanks{e-mail: tsebro@sci.lebedev.ru},
O.E.~Omel'yanovskii,\\ {\em P. N. Lebedev Physics Institute,
Russian Academy of Sciences,}\\ {\em II7924 Moscow, Russia;}\\
{\em International Laboratory of High Magnetic Field and Low
Temperatures,}\\ {\em 53421 Wroclaw, Poland}\\ A.P.~Moravskii\\
{\em Institute of Chemical Physics, Russian Academy of
Sciences,}\\ {\em 142432 Chernogolovka, Moscow Region, Russia} }

\date{}

\maketitle

\begin{center}
\begin{minipage}{12cm}
It is found that the magnetization curves of samples of fragments
of cathode carbon deposits with a high content of multiwalled
nanotubes exhibit a pronounced irreversible character, attesting
to the induction of persistent currents in the samples and to
magnetic flux trapping, as happens in a multiply connected
superconducting structure. A decrease of the trapped flux in time
could not be observed at low (helium) temperatures with a
measurement time of about 20 h. For intermediate ($\sim 30~$K) and
room temperatures the trapped magnetic flux decays slowly with
characteristic relaxation times of the order of 150 and 15~h,
respectively.

\vspace{12pt}


\end{minipage}
\end{center}

{\bf 1.} The electronic properties of carbon nanotubes are the
subject of great interest and intensive investigations
\cite{Ebbesen:PTD-49-6-26,Smalley:RMP-69-723}. Notable among
recent works are experimental and theoretical works devoted to
coherent electron transport in single-walled nanotubes
\cite{Tans:NAT-386-474,Bockrath:SCI-275-1922,White:NAT-393-240,%
Egger:CM/9803128,Bockrath:NAT-397-598} and theoretical works
\cite{Haddon:NAT-388-31,Lin:PRB-57-6731,Odintsov:CM/9805164},
examining the associated question of circulating, persistent
currents in closed toroidal nanotubes.

Transport spectroscopy data
\cite{Tans:NAT-386-474,Bockrath:SCI-275-1922} show that coherent
electron transport occurs in single-walled nanotubes at very low
temperatures, and it occurs over very large distances, estimated
in Ref.\cite{Tans:NAT-386-474} to be right up to the total
nanotube lengths of several microns. According to the theoretical
results obtained in Ref.\cite{White:NAT-393-240}, the conduction
electrons in a single-walled nanotube are affected by the disorder
averaged over the circumference of the nanotube, and this results
in an increase of the electron mean-free path length with
increasing nanotube diameter and, in consequence, exceptional
ballistic transport properties over unprecedentedly long distances
of the order of 10~$\mu$m and larger, which is what explains the
experimental results. We note that for understandable reasons most
theoretical and experimental works on coherent transport concern
single-walled nanotubes. For this reason, the result obtained in
Ref.\cite{Frank:SCI-280-1744} is very noteworthy. It was shown
there that the conductance of multiwalled carbon nanotubes,
ranging in diameter from 5 to 25 nm and up to 10~$\mu$m long,
measured at room temperature, is quantized in the sense that it
does not depend on the nanotube length or diameter, being equal to
$G_0 = 2e^2/h = (12,9~k\Omega)^{-1}$. According to
Ref.\cite{Frank:SCI-280-1744}, multiwalled carbon nanotubes are
capable of carrying at room temperature a current density above
10$^7$~A/cm$^2$, which indicates that high-temperature electric
transport in such nanotubes is ballistic and occurs without the
release of heat.

On this basis, a very important question is the experimental
observation of persistent currents in closed nanotube structures.
In the present letter we report the results of measurements of
magnetization curves of samples of fragments of cathodic carbon
deposits, formed during arc synthesis of multiwalled nanotubes. It
follows from the data obtained that the carbon medium of such
deposits, which consist of contiguous components with different
morphology, the main one being the multiwalled nanotubes, is
capable of carrying persistent currents at low (liquid-helium)
temperatures or very weakly decaying currents even at high (room)
temperatures, and the magnetic flux in such a nanotube is trapped,
resulting in hysteresis of the magnetization curves, as happens in
a multiply connected superconducting structure.

{\bf 2.} Our experimental samples consisted of small fragments,
extracted from the central part of the carbon deposits formed on
the cathode during arc synthesis of multiwalled nanotubes in the
narrow traditional technology used to fabricate such tubes (see,
for example, Ref.\cite{Ebbesen:PTD-49-6-26}). Ordinarily, such
cathode deposits are subjected to special intense treatment
(ultrasonic dispersing followed by treatment with strong
oxidizers) in order to remove nanoparticles and other carbon
formations and to obtain material consisting of essentially
multiwalled nanotubes only. We used for our investigations
fragments of deposits that were not subjected, after completion of
the arc discharge, to any special treatment that destroys the
structure of the material.

The qualitative composition and internal structure of the deposits
prepared in different technological regimes, including the samples
which we prepared for magnetic measurements, have been
investigated in detail in Ref.\cite{Kiselev:CBN-37-1093}. As a
rule, their central part possesses a columnar structure, oriented
along the growth axis of the deposit. According to
Ref.\cite{Kiselev:CBN-37-1093}, the carbon columns of such a
structure consist of three basic components: multiwalled nanotubes
ranging in diameter from 5 to 45~nm (the most likely value is
$\sim 15$~nm), multiwalled polyhedral particles ranging in size
from 20 to 90~nm, and curved graphitized formations. Their
relative amounts and characteristic sizes are determined by the
parameters of the arc process. While all three components are
present inside the columns, the outer shell of the columns
consists predominantly of only intertwined multiwalled nanotubes.
Multiwalled nanotubes in the form of chaotic braids are also
present in the space between the columns. Nanotubes from different
parts of a deposit are oriented predominantly at obtuse angles
with respect to its growth axis.

\begin{figure}[h]
\begin{center}
\includegraphics[width=12.18cm,height=8.128cm]{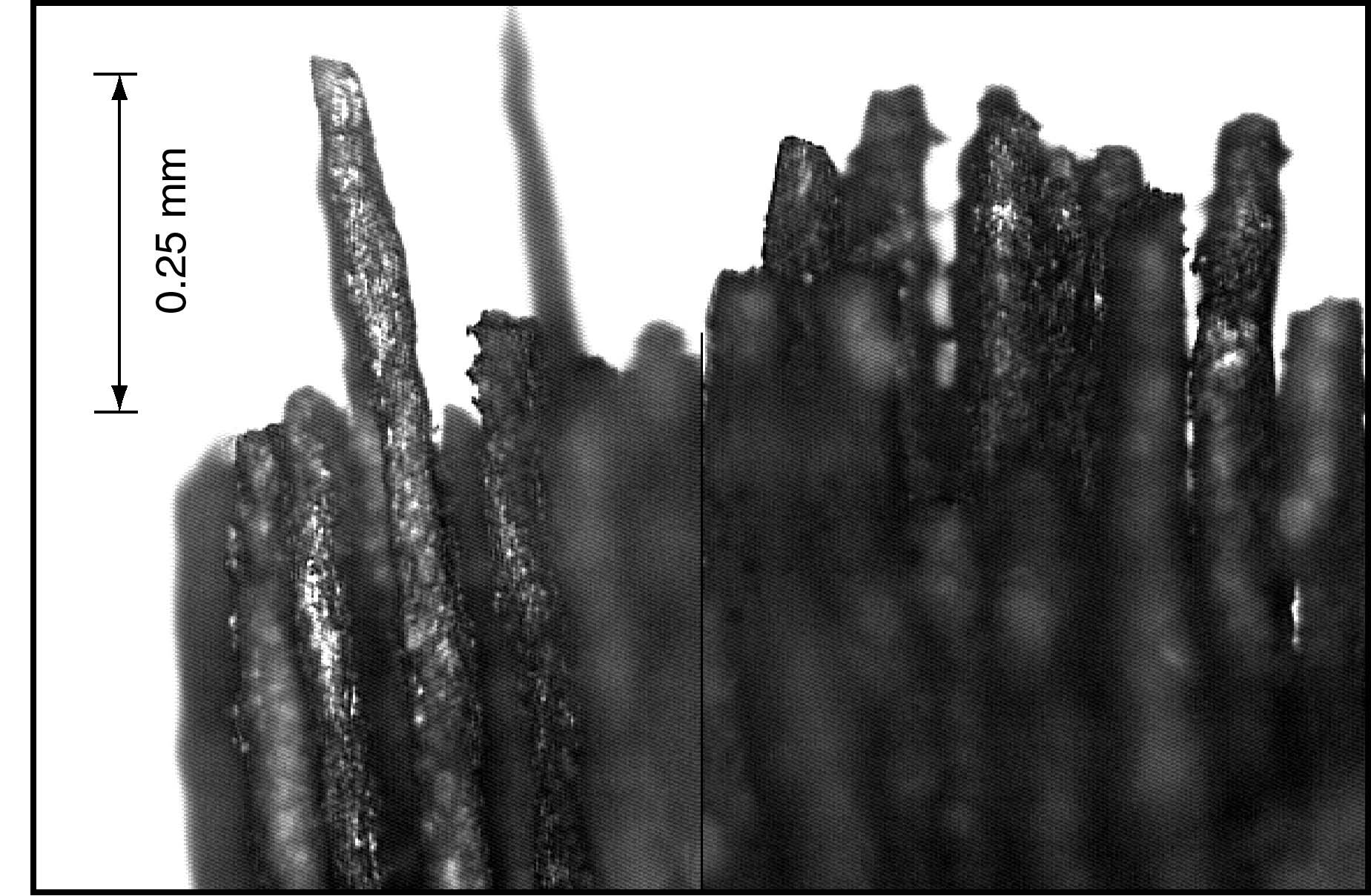}
  \caption{Optical image of the profile of the end of sample No.~196-1s,
  consisting of carbon columns assembled together.}\label{fig1}
\end{center}
\end{figure}

The samples used for the measurements of the magnetization curves
consisted of either (1) carbon columns (average column diameter
$\sim$~50~$\mu$m) obtained from the center of the deposit,
assembled and held together with a negligible amount of
nonmagnetic glue, and oriented along the $z_c$ axis or (2) small,
$\sim$~2.5~mm in diameter, bulk cylinders cut from the center of
the deposit along the growth axis $z_d$. In the first variant the
samples ranged from one to several milligrams in mass, and they
were used for measurements of the magnetization curves in weak
magnetic fields ($< 500$~Oe) using a SQUID magnetometer with a
sensitivity of the order of $5\cdot\,10^{-9}$~emu with respect to
the magnetic moment. Figure~\ref{fig1} shows an optical image of
the profile of the end of a sample (No.~196-1s), consisting of
carbon columns assembled together. In the second variant the
cylindrical samples were of the order of several tens of
milligrams in mass, and they were used for magnetic measurements
in strong magnetic fields using a self-compensated magnetometer
with a capacitance sensor \cite{Wroclaw:SPEC}.

\begin{figure}[h]
\begin{center}
\includegraphics[angle=270,width=14cm]{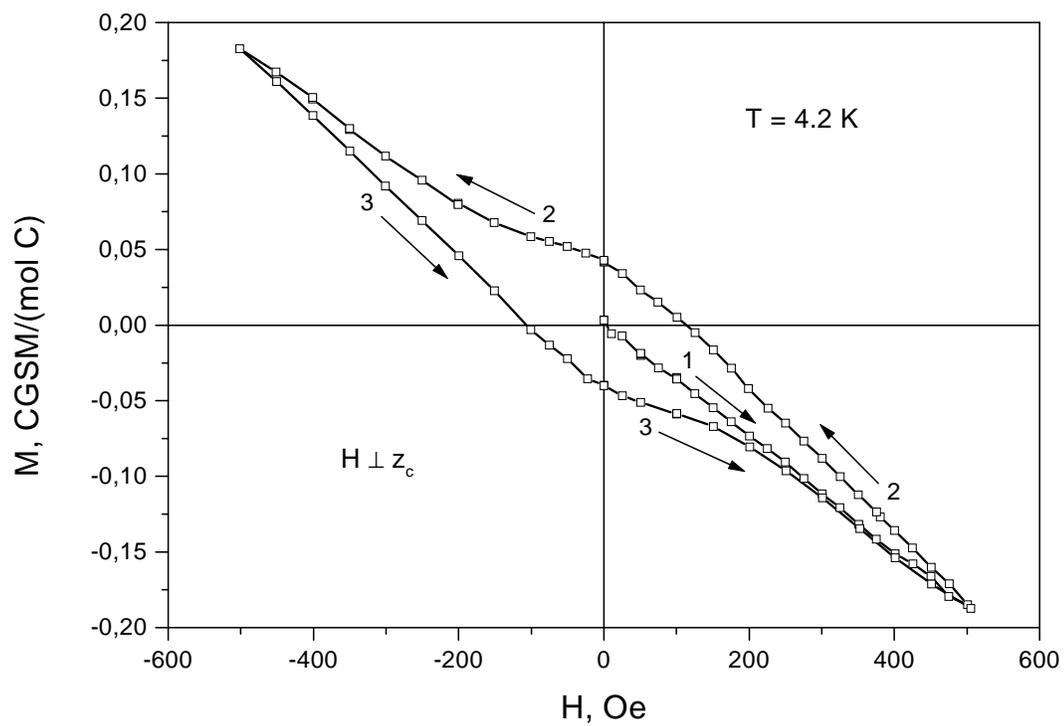}
  \caption{Hysteresis loop of the magnetization curve of sample No.~196-1s
  at $T = 4.2$~K. The sample mass is 1.65~mg. The magnetic field is directed
  perpendicular to the axis of the carbon columns.}\label{fig2}
\end{center}
\end{figure}

{\bf 3.} \emph{Weak magnetic fields.} Figure~\ref{fig2} shows the
results of measurements performed on a SQUID magnetometer at $T =
4.2$~K of the magnetization curve of sample No.~196-1s (sample
mass 1.65~mg, $H\ \bot\ z_c$), held for a long time at room
temperature in a zero (the Earth's) magnetic field. The initial
increase of the field in this case always results in a virtually
linear dependence $M(H)$ (curve~1) with slope (magnetic
susceptibility) for this sample equal to $\chi = -3.8\cdot
10^{-4}$~emu/(mole~C). As the magnetic field decreases (curve~2),
trapped magnetic flux, corresponding in this case to a
paramagnetic moment $M_r \approx 0.04$~emu/(mole~C), remains in
the sample, and with further cycling of the magnetic field from
-500 to +500~Oe (curves~2 and~3) a characteristic hysteresis loop
is observed.

Long-time (up to 20~h) observations of $M_r$, at liquid-helium
temperature did not show, within the limits of measurement
accuracy $(\sim 1\%)$, any appreciable decrease of the moment;
this indicates that the currents induced in the sample are
persistent at low (liquid-helium) temperatures. Measurements of
$M_r$ is a function of time and temperature at higher temperatures
showed that up to room temperatures $M_r$ depends mainly on the
time and not the temperature. Thus, when the sample is heated to
intermediate temperatures ($\sim 20$~K), $M_r$ does not change but
it already shows appreciable, exponential, relaxation with
characteristic relaxation time $\tau_0 \sim 150$~h. When the
sample is heated relatively rapidly (in order to eliminate the
time factor) up to room temperature, $M_r$ decreases by several
percent, and measurements of the time dependencies $M_r(t)$ showed
that for such high temperatures the relaxation time $\tau_0$
remains quite long, of the order of 15~h. Therefore, in order to
return the sample essentially into the initial state (which means,
for example, $M_r$ is decreased to a level $< 1\%$ of the initial
value), the sample must be held at room temperature for
approximately three days.

A repeated check of different samples prepared from different
parts of the same deposit and from different deposits showed that
hysteresis of the magnetization curves occurs in virtually all
cases --- only its magnitude changes, and in very wide limits from
sample to sample. For a number of samples the contribution of the
irreversible part of the magnetization was very small, and the
hysteresis properties of the $M(H)$ curves could be illustrated
satisfactorily only by presenting the difference curves
$M(H)-\chi_0H$ (where $\chi_0$ is the static magnetic
susceptibility at the extreme points of the hysteresis loop). In
this sense the data presented in Fig.~\ref{fig2} are better, in
terms of the magnitude of the effect, than the data that we
obtained in weak magnetic fields for samples consisting of carbon
columns.

\begin{figure}[h]
\begin{center}
\includegraphics[angle=270,width=14cm]{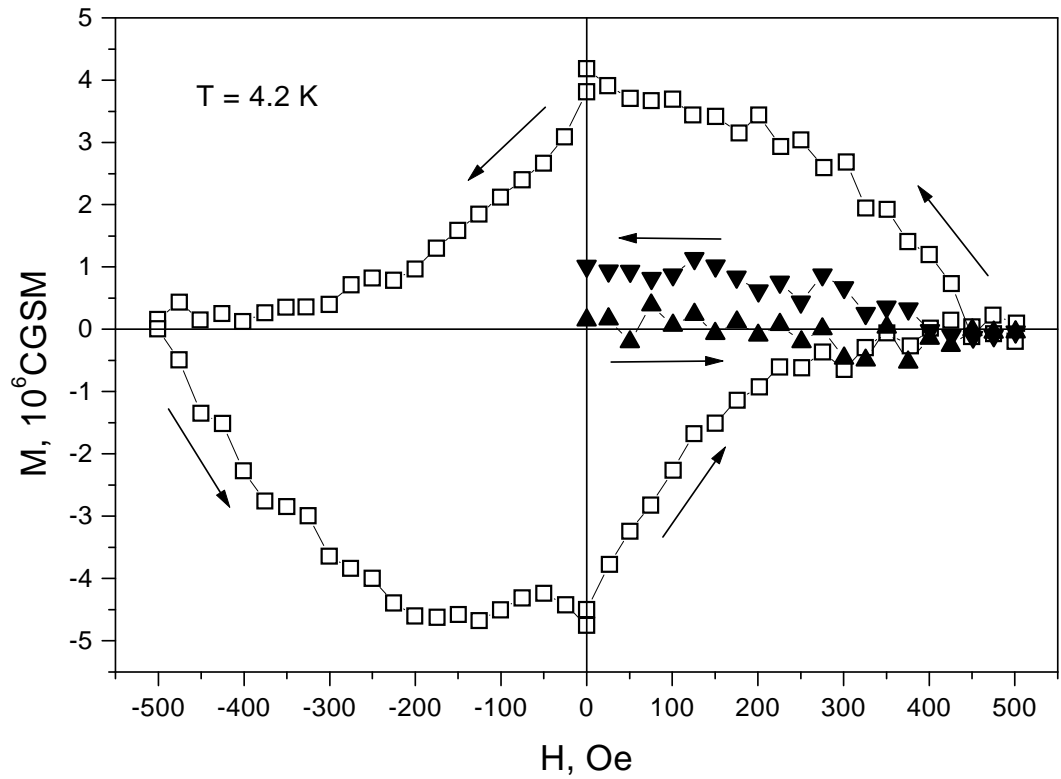}
  \caption{Hysteresis anisotropy of the magnetization curves for sample
  No.~196-11 at $T = 4.2$~K. The absolute values of the magnetic moment
  minus $\chi_0H$ are presented (see text):
  $\square$ -- complete hysteresis loop for $H \perp z_c$,
  initially increasing $(\blacktriangle)$
  and decreasing $(\blacktriangledown)$ magnetic field for $H\ \|\ z_c$.}
  \label{fig3}
\end{center}
\end{figure}

We also note that the magnitude of the hysteresis of the
magnetization curves of samples consisting of carbon columns
depends very strongly on the direction of the magnetic field
relative to the axis $z_c$ of the columns. The magnetic
susceptibility itself or the magnetization of the sample is also
anisotropic. But the magnetization anisotropy in weak magnetic
fields is very small --- of the order of several percent --- and
the magnetic susceptibility is, as a rule, larger for $H\ \|\
z_c$, whereas the hysteresis of the $M(H)$ curves changes
severalfold as a function of the orientation of the sample, the
effect being maximum for $H \perp z_c$. As an example,
Fig.~\ref{fig3} shows for one of the samples (No.~196-11) the
complete hysteresis loop of the form $M(H)-\chi_0H$ for $H \perp
z_c$ and the initial curves $M(H)-\chi_0H$ with increasing and
decreasing magnetic field with the orientation $H\ \|\ z_c$. It is
evident that the trapped flux $M_r$ for $H\ \|\ z_c$ is
approximately four times smaller than for $H \perp z_c$.

\begin{figure}[h]
\begin{center}
\includegraphics[angle=270,width=14cm]{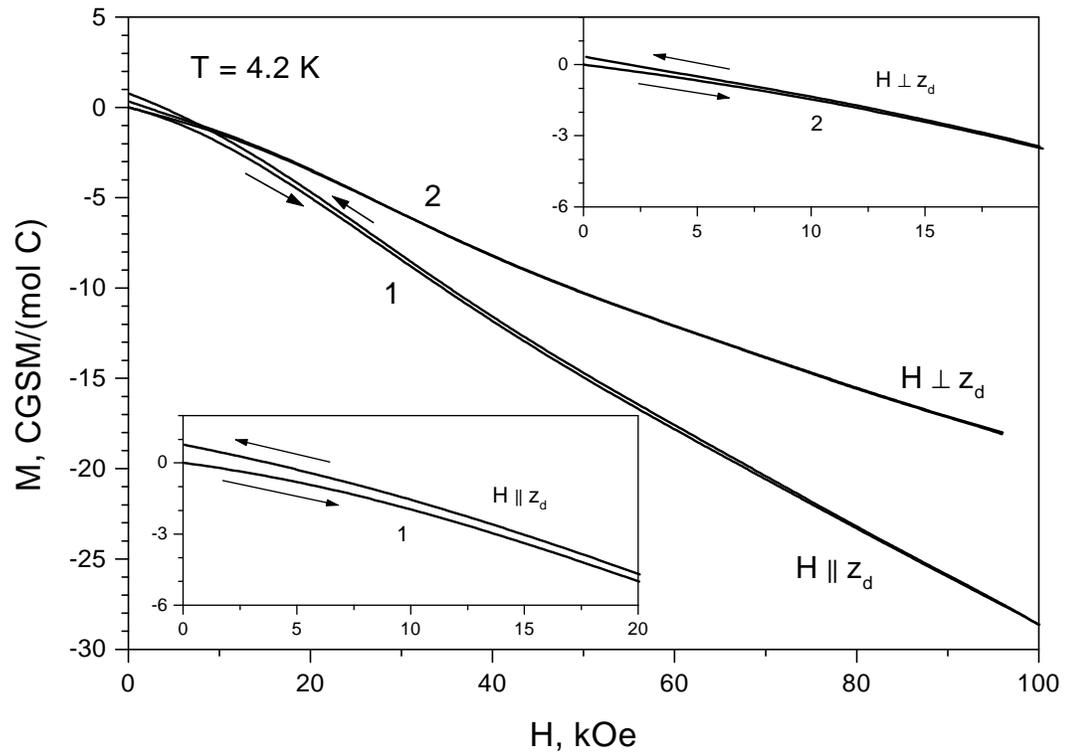}
  \caption{Magnetization curves for sample No. 140 in strong,
  increasing and decreasing magnetic fields with orientation
  along (1) and perpendicular (2) to the growth axis of
  the deposit. $T = 4.2$~K. The sample mass is 66.6 mg.
  Inserts: Initial sections of the curves for different orientations
  of the magnetic field.}\label{fig4}
\end{center}
\end{figure}

{\bf 4.} \emph{Strong magnetic fields.} The measurements were
performed at $T = 4.2$~K in the nonuniform field of a
superconducting solenoid using a self-balancing magnetometer with
a capacitance sensor \cite{Wroclaw:SPEC} on cylindrical samples
cut from the central part of the deposit along the growth axis
$z_d$ of the deposit. The magnetic field was not switched, i.e.,
the $M(H)$ curves were measured with increasing and decreasing
fields. Figure~\ref{fig4} shows magnetization curves for one of
the samples (No.~140) with increasing and decreasing magnetic
field for two orientations: $H\ \|\ z_d$ (curve~1) and $H \perp
z_d$ (curve~2). These data illustrate the general pattern of the
results obtained in strong magnetic fields. In the first place,
the magnetization curves are strongly nonlinear, which indicates
the complicated character of the magnetic interactions in the
system. We note that in the process there is a strongly nonlinear
field dependence of the magnetization anisotropy, where rapid
growth of the ratio $M_{\|}/M_{\bot}$ in the field range 0--20~kOe
is replaced by slow monotonic growth of the ratio up to
$M_{\|}/M_{\bot} \sim 1.5$ for $H \sim 100$~kOe. In the second
place, the magnetization curves are irreversible even in strong
magnetic fields. Just as in weak magnetic fields and for carbon
column samples, the hysteresis of the $M(H)$ curves is appreciably
anisotropic, but the hysteresis is greater for$H\ \|\ z_d$ (see
Fig.~\ref{fig4}). We also note that when the field completely
leaves the region of strong magnetic fields, for some samples the
residual moment $M_r$ reaches values $\sim 2$~emu/(mole~C) for $H\
\|\ z_d$.

{\bf 5.} In summary, we have observed that samples of fragments of
cathode carbon deposits, which were not damaged by special
treatment in order to remove the multiwalled nanotubes contained
in them, can carry persistent magnetic-field induced currents at
low temperatures (liquid-helium) or very weakly decaying currents
at high (room) temperatures. This property is observed for
magnetization curves that show a pronounced irreversible
character, i.e., magnetic flux is trapped in the samples, as
happens in a multiply connected superconducting structure.

At present one can only surmise how the system of paths conducting
persistent or weakly decaying currents in the carbon medium of
such samples is organized. It is possible that the structure is
similar to a so-called ``Mendelssohn sponge''
\cite{Mendelssohn:PRS-A152-34} (a multiply connected system of
thin superconductor strands in a normal matrix), and the character
of the irreversible behavior of its magnetization corresponds to
the well-known critical-state model \cite{Bean:PRL-8-250}. If this
is so, then the virtually linear dependence $M(H)$ with the field
increasing initially (see curve~1 in Fig.~\ref{fig2}) indicates
that the critical current of the filaments is very high, and the
corresponding field-dependent penetration depth of the magnetic
field in such a sponge is small. What comprises such a sponge
structure is also unclear, but in this case there is a system of
interconnected nanotubes in which electric transport is loss-free
or the losses are negligibly small. According to
electron-microscope data \cite{Kiselev:CBN-37-1093}, the structure
of the outer shell of the carbon columns of the deposit matches
such a sponge structure; such a shell consists of a quite dense
network of intertwined and interconnected nanotubes. Since the
planes of the cells of such a network are perpendicular to the
axis of the columns, the large anisotropy of the hysteresis of the
magnetization curves and the fact that the trapped flux is greater
in the case $H \perp z_c$ become understandable. A more sparse but
also micron-size network of nanotubes is also present in the space
between the columns along their entire length
\cite{Kiselev:CBN-37-1093}. This reticular structure, where the
planes of the cells are perpendicular to the growth axis of the
deposit, likewise seems to trap magnetic flux well, and its effect
becomes determining in bulk samples cut from a deposit, the
trapped flux being maximum for $H\ \|\ z_d$. Since, as noted
above, the hysteresis of the $M(H)$ curves varies very strongly
from sample to sample, it can be inferred that it is the quality
of the intertube connections that determines the wholeness of the
sponge and the corresponding trapping of the magnetic flux,
leading to hysteresis of the magnetization curves. As a rule,
appreciable hysteresis of the magnetization curves has always been
observed in samples whose magnetic susceptibility at low
temperatures was much greater in absolute magnitude than the value
adopted for nanotubes $\chi \sim -3\cdot 10^{-4}$~emu/(mole~C)
\cite{Haddon:NAT-378-249,Dresselhaus:SFCN-1996}. Apparently,
persistent currents already make a large contribution to the
diamagnetic response of such samples to an external magnetic
field. It has not been ruled out that the proposed
current-carrying sponge does not work in the manner that has been
supposed, as a whole, in which case it must be assumed that
magnetic flux is trapped in individual cells of the nanotube
network which are unconnected or weakly connected with one
another. Further investigations are required to clarify the nature
of the observed persistent currents and the corresponding
current-carrying structure.

We thank the State Scientific and Technical Program ``Topical
Problems in Condensed-Matter Physics'' for support (``Fullerenes
and Atomic Clusters'' No.~2-5-99).

\newpage
\pagestyle{empty}
\mbox{} \vspace{5cm}

\newpage
\mbox{} \vspace{5cm}

\newpage
\mbox{} \vspace{5cm}

\newpage
\mbox{} \vspace{5cm}

\end{document}